\documentclass[aps,prd,preprint,notitlepage,showkeys,superscriptaddress,nofootinbib,groupedaddress]{revtex4-2}

\usepackage[utf8]{inputenc}
\usepackage[T1]{fontenc}
\usepackage[english]{babel}
\usepackage{amsmath,amssymb}
\usepackage{bm}
\usepackage{graphicx}
\usepackage{float}
\usepackage{enumitem}
\usepackage{microtype}
\usepackage{placeins}
\usepackage{hyperref}

\hypersetup{
    colorlinks=true,
    linkcolor=blue,
    citecolor=blue,
    urlcolor=blue,
    pdftitle={Simulating the emission source function in very peripheral relativistic heavy-ion collisions},
    pdfauthor={Angel Reina Ram{\'i}rez, Volodymyr Magas, Juan M. Torres-Rincon}
}

\makeatletter
\def\frontmatter@abstract@produce{%
  \par
  \addvspace{\frontmatter@preabstractspace}%
  \begingroup
    \dimen@\baselineskip
    \setbox\z@\vtop{\unvcopy\absbox}%
    \advance\dimen@-\ht\z@\advance\dimen@-\prevdepth
    \@ifdim{\dimen@>\z@}{\vskip\dimen@}{}%
  \endgroup
  \begingroup
    \prep@absbox
    \unvbox\absbox
    \post@absbox
  \endgroup
  \@ifx{\@empty\mini@notes}{}{\mini@notes\par}%
  \addvspace\frontmatter@postabstractspace
}%
\def\frontmatter@abstractheading{%
 \begingroup
  \centering\large\bfseries
  \abstractname
  \par
 \endgroup
}%
\def\@keys@name{\textbf{Keywords:} }%
\makeatother

\begin{document}
	
		\title{Simulating the emission source function \\ in very peripheral relativistic heavy-ion collisions}
		\author{Angel Reina Ram{\'i}rez}
		\author{Volodymyr Magas}
	       \author{Juan M. Torres-Rincon}
        
		\affiliation{
			Departament de F\'isica Qu\`antica i Astrof\'isica,
			Universitat de Barcelona,
			Mart\'i i Franqu\`es 1, 08028 Barcelona, Spain}%
		\affiliation{
			Institut de Ci\`encies del Cosmos (ICCUB),
			Universitat de Barcelona,
			Mart\'i i Franqu\`es 1, 08028 Barcelona, Spain}%
	
		\date[]{}

\begin{abstract}
We present a microscopic femtoscopic investigation of the space-time emission source in low-multiplicity (80--90\% centrality) $\text{Pb}-\text{Pb}$ collisions at $\sqrt{s_{NN}} = 5.02~\text{TeV}$, using a hybrid transport framework coupling SMASH initial conditions, 3+1D viscous hydrodynamics (vHLLE), and a SMASH hadronic cascade afterburner. Access to the full space-time evolution of the produced particles allows for the direct reconstruction of emission source functions in both the Longitudinally Comoving System and the Pair Rest Frame. We study identical pairs of mesons ($\pi\pi$, $KK$) and baryons ($pp$), as well as cross-species combinations, while explicitly isolating the contributions from primordial particles, hadronic rescattering, and long-lived resonance decays. Our analysis shows that  simple Gaussian fits are inadequate for precision source analyses, whereas a combined Gaussian plus exponential parametrization better captures the compact core and the extended tail. Finally, the extracted Gaussian core radii, $R_{\text{core}}$, for our very peripheral collisions demonstrate an approximate $m_T$-scaling behavior rather similar to the one observed in pp reactions.
\end{abstract}

		
		\maketitle
	
	\section{Introduction}
	
Relativistic heavy-ion collisions create short-lived systems of strongly-interacting matter whose evolution must be inferred from the observed final-state particles. Momentum-space observables such as particle spectra and anisotropic flow constrain the collective expansion of the system, but they do not, by themselves, determine the space--time structure of particle emission. Femtoscopy (often referred to as HBT interferometry, named after the Hanbury--Brown and Twiss effect in astronomy~\cite{HanburyBrown:1956bqd})  provides the complementary information needed to probe the size, shape, and temporal structure of the emitting source~\cite{Wiedemann:1999qn,Lisa:2005dd}.
It is one of the few techniques that allows us to directly infer the spatial and temporal dimensions of the  emitting source on the scale of femtometers  and zeptoseconds ($\sim 10^{-21}\,\mathrm{s}$) \cite{Wiedemann:1999qn,Heinz:1999rw,Lisa:2005dd}.

Femtoscopy relies on quantum statistics and final-state interactions (FSI) between pairs of particles emitted close to each other in momentum space. 
Classical HBT interferometry ~\cite{HanburyBrown:1956bqd, Wiedemann:1999qn,Heinz:1999rw} deals with pairs of  identical bosons. When two identical bosons (e.g., pions) are produced in a relativistic heavy-ion collision, their overall wave function must be symmetric under particle exchange. This leads to quantum mechanical Bose-Einstein correlations: identical bosons produced close together in phase space tend to ``bunch'' together. 
Modern femtoscopy also deals with non-identical hadron-hadron pairs (e.g., $p-\Lambda$~\cite{ALICE:2018ysd}, $p-\Xi$~\cite{ALICE:2019hdt}, $K-p$~\cite{ALICE:2019gcn}, hyperon-hyperon~\cite{ALICE:2018ysd,ALICE:2019eol}), where quantum statistics play a lesser or no role, but FSI dominate the small relative momentum region~\cite{Heinz:1999rw,Lednicky:2005tb}. This has turned femtoscopy into a powerful tool to measure, for example, hyperon-nucleon and hyperon-hyperon interaction potentials, crucial for understanding neutron star interior physics~\cite{Fabbietti:2020bfg}.

Experimentally, one measures the two-particle correlation function ($\bm{q}$ is the relative 3-momentum), defined as:
\begin{equation}
C(\bm{q}) = \frac{N_{\text{same}}(\bm{q})}{N_{\text{mixed}}(\bm{q})} \ ,
\end{equation}
where $N_{\text{same}}(\bm{q})$ is the distribution of pairs taken from the same event, and $N_{\text{mixed}}(\bm{q})$ is a reference distribution constructed by pairing particles from different events (thereby removing quantum correlations and FSI).

Via the well-known Koonin-Pratt formula~\cite{Pratt:1986cc} this correlation function can be related to the spatial distribution $S(\bm{r})$ of the emitting source,
\begin{equation}
C(\bm{q})  =\int d^3r \, S(\bm{r}) \left| \psi_{\text{rel}}(\bm{q}, \bm{r}) \right|^2 \ ,
\label{C_int_S}
\end{equation}
where $\psi_{\text{rel}}(\bm{q},\bm{r})$ is the relative pair wave function (including quantum statistics, Coulomb repulsion/attraction, and strong interaction potentials) and $r$ is the relative 3-coordinate variable in the center-of-mass frame of the pair.

The source function is the main object of our study. Experimentally, one needs to assume some ansatz for it and then fit its parameters to the experimental data. However, by performing relativistic heavy-ion collision simulations with a microscopic transport code as an afterburner, one can directly simulate the source function~\cite{Sinyukov:2002if,Karpenko:2010te,Shapoval:2013jca,Shapoval:2013bga,Kisiel:2014upa,Sinyukov:2015kga,Kincses:2022eqq,Korodi:2022ohn,Csanad:2024hva,Wang:2024bpl,Nzabahimana:2025ivc,Zhang:2025yqq,Kincses:2025izu,Kisiel:2025jbg,Molnar:2026vgi}. 

Today, several groups that perform heavy-ion collision simulations successfully use the so-called hybrid approach, where different stages of the evolution of the system created in such a collision are simulated with the most suitable approach. Commonly, we can identify three different generic stages: an initial stage or pre-equilibrium state, an intermediate stage---typically simulated with relativistic hydrodynamics---and a final stage, which can be simulated with a transport hadron model, such as Simulating Many Accelerated Strongly-interacting Hadrons (SMASH)~\cite{Weil:2016zrk}.

We use a SMASH + vHLLE + SMASH(afterburner) hybrid model, which has been shown to be efficient at RHIC and LHC~\cite{Schafer:2021csj,Garcia-Montero:2021haa,Gotz:2025wnv,Constantin:2026hwh}. To our knowledge, this is the first application of a hybrid model with SMASH afterburner to source function simulations.

Using SMASH for the  afterburner calculations offers a useful bridge between theory and femtoscopic measurements because they provide direct access to the final four-momenta and emission four-coordinates of the produced particles. This additional information makes it possible not only to reconstruct two-particle correlation functions but also to study their dependence on the different origins of the emitted particles. In this work, we describe and apply a femtoscopic analysis framework designed for afterburner event records in very peripheral $\text{Pb}-\text{Pb}$ collisions. 
Their low final-state multiplicity motivates a comparison with femtoscopic measurements in pp collisions.

The paper is organized as follows. In Sec.~\ref{sec:smash} we first describe the modified version of the SMASH--vHLLE--SMASH event generator, the reconstruction of emission coordinates from the afterburner record, the change from the longitudinally comoving system (LCMS) to the pair rest frame (PRF), and all the observables used in our analysis. In Sec.~\ref{sec:results} we then present the resulting out--side--long and radial source distributions, together with the transverse-mass dependence of the extracted core radius. Finally, in Sec.~\ref{sec:conclusions} we summarize our main conclusions.

\section{Theoretical and computational framework~\label{sec:smash}}

The events are generated within the SMASH--vHLLE--SMASH hybrid model~\cite{Schafer:2021csj}, which couples microscopic initial conditions to the hydrodynamic evolution of the dense medium and subsequently to a final hadronic cascade. Specifically, we use a modified version of the \(3+1\)-dimensional viscous hydrodynamic code vHLLE~\cite{Karpenko:2013wva,vHLLE:software}, dubbed vHLLE-openMP~\cite{vHLLE-openMP:software}---a parallelized implementation that distributes grid computations and freeze-out routines across multiple CPU threads using OpenMP, while preserving the original numerical formulation and physics.

In this evolution chain, SMASH v3.2.2~\cite{Schafer:2021csj,Weil:2016zrk,SMASH:software}---with string excitation and fragmentation handled by PYTHIA 8.315~\cite{Bierlich:2022pfr,PYTHIA}---is first run as a microscopic initial-stage event generator up to a hypersurface of constant proper time \(\tau_0\). Particles crossing this iso-\(\tau\) hypersurface are used to generate an initial state for further hydrodynamical evolution; for this purpose their energy-momentum and conserved-charge densities are converted into the hydrodynamic energy-momentum tensor and charge flows through a fluidization procedure. The resulting medium then evolves with vHLLE-openMP until the so-called particlization hypersurface, defined by a constant critical energy density criterion \(e_{\mathrm{crit}}\). The hypersurface elements and their normal four-vectors are obtained with the Cornelius algorithm~\cite{Huovinen:2012is}. The SMASH hadron-sampler then applies the Cooper--Frye prescription~\cite{Cooper:1974mv} to sample hadrons from these elements~\cite{Karpenko:2015xea,SMASHHadronSampler:software}. The conservation laws during this procedure are satisfied grand-canonically, and thus the same hypersurface elements are sampled many times (\(10^3\) in the present workflow),
producing particle lists that are passed to the final SMASH afterburner~\cite{Schafer:2021csj}.

A central advantage of microscopic afterburner simulations, compared with experimental measurements, is that the space-time history of each particle is directly accessible, allowing us to reconstruct its emission point. For the present analysis, we use the microscopic information stored in the SMASH afterburner output: the final particle coordinates and four-momenta, the time of the last microscopic interaction or decay, and information on the particle origin. This output is analyzed with the public code SMASH Afterburner Analysis~\cite{AfterburnerAnalysis:software}, which implements the source reconstruction, pair construction, frame transformations, histogramming, and fitting procedures described below. Because the final coordinates correspond to particles after their last interaction and subsequent free streaming to the output time, we first need to map each particle back to its emission point. This is done by propagating it along a straight-line trajectory from its final position to the known time of its last interaction or decay, using its final momentum. Direct primordial particles with no subsequent collisions are mapped back by assigning them the coordinates stored in the SMASH-hadron-sampler particle list; in this case, the emission coordinate coincides with the particlization point sampled on the hydrodynamic switching hypersurface.

This procedure gives, for each particle, the four-momentum and reconstructed emission four-coordinates used in the source analysis. Particles are then grouped event by event into specific species samples from which the corresponding pairs are later constructed. 

For each particle $i=\{ 1,2\}$ in a given pair, we denote the four-momentum by \(p_i^\mu=(E_i,\bm{p}_i)\) and the reconstructed emission coordinate by \(x_i^\mu=(t_i,\bm{x}_i)\). Both are given in the laboratory (simulating) frame. We then define the relevant pair quantities: the total pair momentum \(P^\mu\), the average pair momentum \(K^\mu\), the relative momentum \(q^\mu\), and the relative separation of two emission points \(r^\mu\), as
\begin{equation}
\begin{aligned}
P^\mu &= p_1^\mu + p_2^\mu \ ,
&
K^\mu &= \frac{1}{2}(p_1^\mu + p_2^\mu) \ ,
\\
q^\mu &= p_1^\mu - p_2^\mu \ ,
&
r^\mu &= x_1^\mu - x_2^\mu \ .
\end{aligned}
\end{equation}
From the transverse components to the beam axis, we also define the transverse relative separation, pair transverse momentum, and the transverse mass as
\begin{equation}
r_T = \sqrt{r_x^2+r_y^2} \ , 
\qquad
K_T=\sqrt{K_x^2+K_y^2} \ ,
\qquad
m_T=\sqrt{K_T^2+\langle m\rangle^2} \ ,
\end{equation}
where \(\langle m\rangle\) denotes the average particle mass of the pair,
\begin{equation}
\langle m\rangle=\frac{m_1+m_2}{2}.
\end{equation}

The relative four-vector \(r^\mu\) is then transformed to the LCMS, defined by the condition that the longitudinal total momentum (i.e. along the beam axis, $z$) vanishes, 
\(P^z_{\mathrm{LCMS}}=0\). In other words, it is Lorentz boosted along the beam direction with velocity \(\beta_{\mathrm{LCMS}}=P^z/P^0\). 

After the boost to the LCMS, the spatial separation $\bm{r}_{\mathrm{LCMS}}$ 
[$r^\mu_{\mathrm{LCMS}}=(\Delta t_{\mathrm{LCMS}},\bm{r}_{\mathrm{LCMS}})$] 
is projected into the out--side--long (OSL) basis introduced in the Bertsch--Pratt parametrization~\cite{Pratt:1986cc,Bertsch:1988db},
\begin{equation}
\hat e_{\mathrm{out}}=\frac{\bm{K}_T}{|\bm{K}_T|} \ ,
\qquad 
\hat e_{\mathrm{long}}=\hat z \ ,
\qquad
\hat e_{\mathrm{side}}=\hat e_{\mathrm{long}}\times \hat e_{\mathrm{out}} \ ,
\end{equation}
and, correspondingly 
\begin{equation}
r_{\mathrm{out}}=\bm{r}_{\mathrm{LCMS}}\cdot\hat e_{\mathrm{out}} \ ,
\qquad
r_{\mathrm{side}}=\bm{r}_{\mathrm{LCMS}}\cdot\hat e_{\mathrm{side}} \ ,
\qquad
r_{\mathrm{long}}=\bm{r}_{\mathrm{LCMS}}\cdot\hat e_{\mathrm{long}} \ .
\label{eq:osl-components-lcms}
\end{equation}

The OSL basis is conveniently defined in the LCMS, since the transverse pair momentum provides a well-defined ``out'' direction. However, source function analyses are most naturally performed in the PRF, where the center-of-mass motion of the pair is removed and the spatial separation directly characterizes the relative geometry of the two-particle system. A direct OSL decomposition in the PRF would be ambiguous since, by definition, the total three-momentum of the pair vanishes and no transverse pair-momentum direction remains available to define the out axis. We therefore use the LCMS as an intermediate frame: after projecting the LCMS spatial separation onto the OSL basis, the obtained 
$
r_{\mathrm{LCMS}}^\mu
$
is then boosted to the PRF; the corresponding boost velocity is
$
\boldsymbol{\beta}_{\mathrm{PRF}\leftarrow\mathrm{LCMS}}
=\bm{P}_{\mathrm{LCMS}}/P^0_{\mathrm{LCMS}}$.
Since \(P^{z}_{\mathrm{LCMS}}=0\), the boost is purely transverse, i.e. along the out direction, 
\(\boldsymbol{\beta}_{\mathrm{PRF}\leftarrow\mathrm{LCMS}}
 =\beta_T\hat e_{\mathrm{out}}\). The transformation reads
\begin{equation}
\label{eq:lcms-to-prf-components}
\left\{
\begin{aligned}
\Delta t^*
&=
\gamma_T
\left(
\Delta t_{\mathrm{LCMS}}-\beta_T r_{\mathrm{out}}
\right) \ ,
\\
r_{\mathrm{out}}^*
&=
\gamma_T
\left(
r_{\mathrm{out}}-\beta_T\Delta t_{\mathrm{LCMS}}
\right) \ ,
\\
r_{\mathrm{side}}^*
&=
r_{\mathrm{side}} \ ,
\\
r_{\mathrm{long}}^*
& =
r_{\mathrm{long}} \ ,
\end{aligned}
\right.
\end{equation}
where \(\gamma_T=(1-\beta_T^2)^{-1/2}\) and the asterisk denotes the coordinates in the PRF. In this way, the PRF retains the LCMS OSL definition while incorporating the space-time mixing generated by the Lorentz boosts.

Notice that the particles in the pair do not necessarily share the same emission time. Consequently, $r^0=\Delta t \ne 0$ in none of these reference frames (laboratory, LCMS or PRF) for most pairs.  Since experimentally there is no way to know the particle emission time, the $\Delta t$ dependence of the source function is usually integrated over~\cite{Heinz:1999rw,Lisa:2005dd}. 

From the transformed spatial components, Eq. (\ref{eq:lcms-to-prf-components}), we define the modulus of the full radial separation as
\begin{equation}
r_{\mathrm{full}}^{*}
=
\sqrt{
\left(r_{\mathrm{out}}^{*}\right)^2+
\left(r_{\mathrm{side}}^{*}\right)^2+
\left(r_{\mathrm{long}}^{*}\right)^2
} \ .
\label{eq:radial-separation}
\end{equation}

In our simulation the source function for a given hadron pair can be reconstructed  from the output of SMASH in the following way: 
\begin{equation}
S(\bm{r}^{(*)})=\frac{1}{N_{\mathrm{pairs}}} \sum_{j=1}^{N_{\mathrm{pairs}}} \delta(\bm{r}^{(*)}-\bm{r}_j^{(*)}) \ , 
\end{equation}
where $j$ runs over the number of pairs in the system. In practice, however, we evaluate one-dimensional distributions by constructing histograms of the three OSL components and those in the PRF.

To characterize the spatial scales of the source, both one-dimensional OSL distributions and the radial distribution  are fitted with a Gaussian parameterization, 
\begin{equation}
S_{\mathrm{G}}(r_u^*)
=
A\exp\!\left[-\frac{r_u^{*2}}{4R_u^2}\right] \ , \label{eq:gauss}
\end{equation}
and with a Gaussian-plus-exponential (G+E) form,
\begin{equation}
S_{\mathrm{G+E}}(r_u^*)
=
A\left[
f_{\mathrm{core},u}\exp\!\left(-\frac{r_u^{*2}}{4R_{\mathrm{core},u}^2}\right)
+(1-f_{\mathrm{core},u})\exp\!\left(-\frac{|r_u^{*}|}{R_{\mathrm{tail},u}}\right)
\label{eq:gaussexp}
\right] \ ,
\end{equation}
where $A$ is a scale parameter. In Eq.~(\ref{eq:gaussexp}), $f_{\mathrm{core},u}$ is a fitting parameter accompanying the Gaussian (core) term. 
The Gaussian core radius, $R_{\mathrm{core},u}$, in the G+E fit is interpreted as the compact core scale, while the exponential term parametrizes the long-range tail with a characteristic scale $R_{\mathrm{tail},u}$.

Several studies of pp collisions measured by the ALICE Collaboration have analyzed the $m_T$ dependence of Gaussian core radius, $R_{\mathrm{core}}$, for different hadron pairs, mostly those where the strong interaction is negligible (as for same-sign pions~\cite{ALICE:2023sjd}) or well known (as in proton-proton correlations~\cite{ALICE:2020ibs}). They found the so called ``$m_T$-scaling'', i.e. the effect that the $m_T$ dependence of Gaussian core of the source is, in a very good approximation, universal for the different hadron pairs, independently of the species~\cite{Lisa:2005dd}. 
 While $m_T$-scaling has been extensively tested in pp collisions~\cite{Fabbietti:2020bfg}, there are no solid indications that it might hold in heavy-ion collisions, see, for example, 
 a recent analysis by the ALICE group for central and mid-peripheral $\text{Pb}-\text{Pb}$ collisions \cite{ALICE:2025wuy}. 
 
In this work, we decided to focus on very peripheral collisions, since these are the closest to p+p collisions, in terms of the final charged multiplicity. We would like to test whether 
$m_T$-scaling still applies in this system, before considering more central collisions, where flow and collective effects dominate.

\section{Results~\label{sec:results}}

Having defined the source observables considered in this work and the corresponding frame transformations, we now apply the analysis to a sample of \(10^6\) $\text{Pb}-\text{Pb}$ events at \(\sqrt{s_{NN}}=5.02~\mathrm{TeV}\) in the 80--90\% centrality class. Events are generated within the SMASH--vHLLE--SMASH hybrid model~\cite{Schafer:2021csj} with a hydrodynamic initial proper time of \(\tau_0=0.5~\mathrm{fm}/c\), a shear viscosity to entropy density ratio \(\eta/s=0.08\), and a particlization criterion based on a critical energy density \(e_{\mathrm{crit}}= 0.5~\mathrm{GeV}/\mathrm{fm}^3\).

The analysis of the simulated events was performed with SAFA (SMASH Afterburner Femtoscopy Analysis), a dedicated framework developed to process the output of the SMASH--vHLLE simulation chain and reconstruct the femtoscopic source observables considered in this work. SAFA handles the particle-origin classification, pair construction, reference-frame transformations, source-distribution analysis, and the corresponding fits. The code is publicly available in~\cite{SAFA:software}.

For all particle species, we impose a cut on pseudorapidity \(|\eta|<0.8\) from the ALICE detector, together with transverse-momentum cuts which depend on the species: for pions and kaons we use \(0.14<p_T<4.0~\mathrm{GeV}/c\) and \(0.4<p_T<1.4~\mathrm{GeV}/c\), respectively, following Ref.~\cite{ALICE:2023sjd}. For nucleons we use \(0.5<p_T<4.05~\mathrm{GeV}/c\), following Ref.~\cite{ALICE:2019pSigma}.

For all of the observables which we show, we select particles from three different origins: 
\begin{enumerate}
    \item Primordial particles: those emitted directly from the particlization hypersurface and suffered no interaction in their evolution.
    \item Primordial-plus-rescattered particles: accounts from the primordial ones plus those whose last emission point follows from hadronic rescattering.
    \item All origins: all generated hadrons, that is, primordial-plus-rescattered particles plus particles emitted from resonance decays.
\end{enumerate} 

Such a separation allows us to study the effects of hadronic rescattering and resonance decays in the source.

\subsection{Emission-coordinate distributions}
\label{FO_distribution}

\begin{figure}[pt]
\centering
\includegraphics[width=\linewidth]{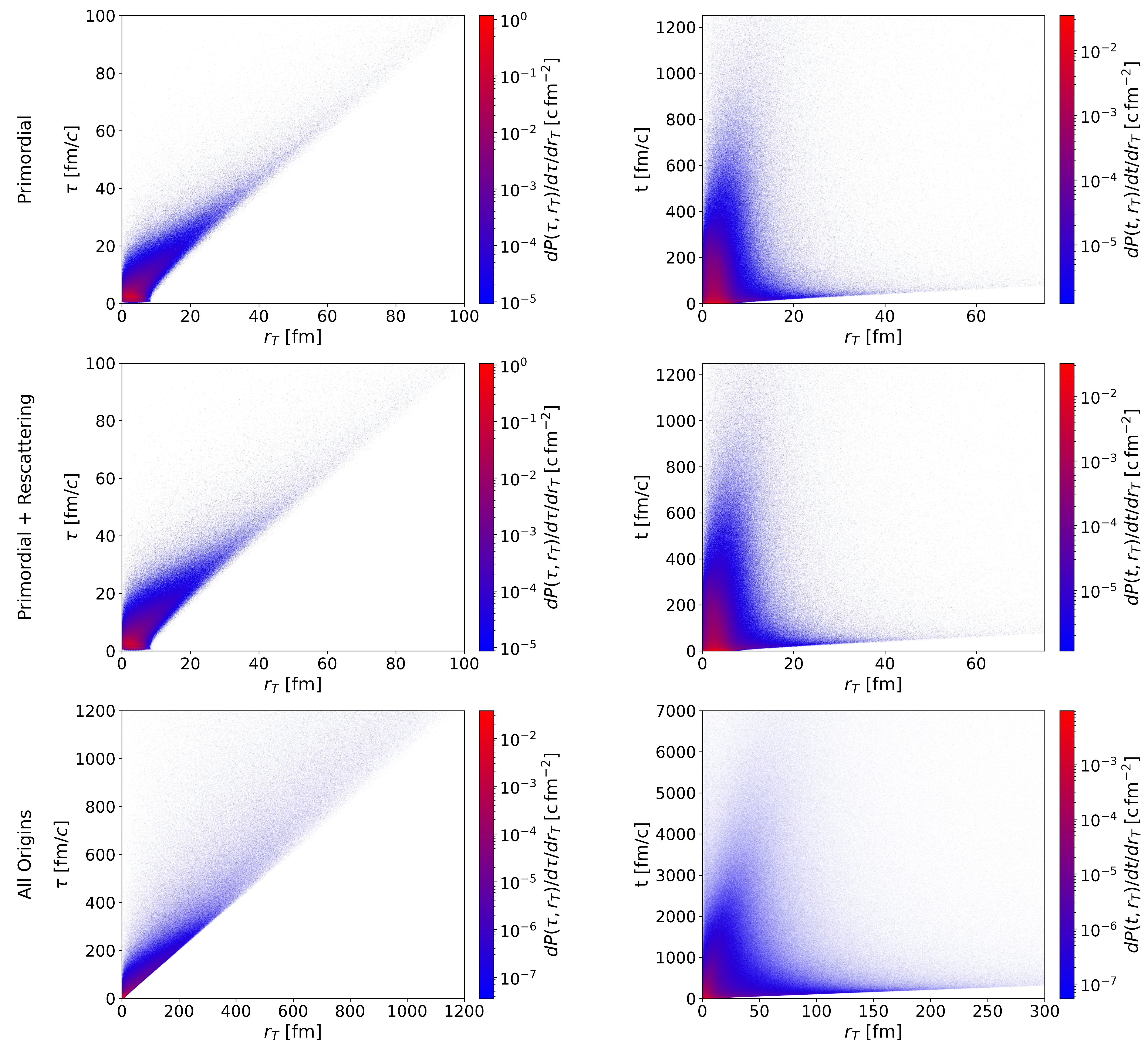}
\begingroup
\caption{Emission-coordinate distributions of individual particles in the laboratory frame projected onto the \(\tau\)--\(r_T\) and \(t\)--\(r_T\) planes. All produced particle species are included except photons. The rows correspond to primordial particles, primordial-plus- rescattered particles, and hadrons from all origins, respectively.}
\label{fig:emission-coordinates}
\endgroup
\end{figure}

\begin{figure}[pt]
\centering
\includegraphics[width=\linewidth]{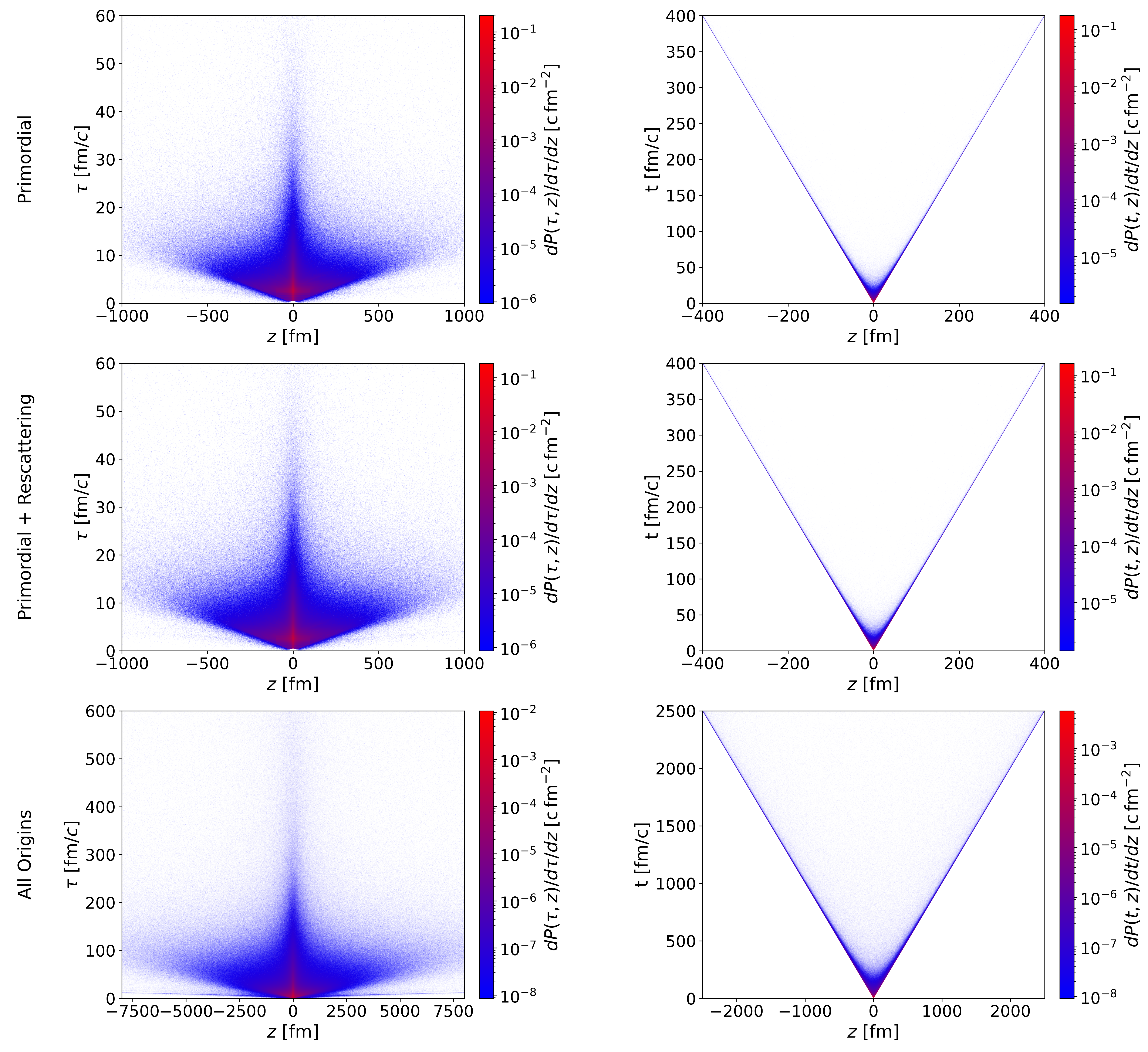}
\begingroup
\caption{Emission-coordinate distributions of individual particles in the laboratory frame projected onto the \(\tau\)--\(z\) and \(t\)--\(z\) planes. All produced particle species are included except photons. The rows correspond to primordial particles, primordial-plus-rescattered particles, and hadrons from all origins, respectively.}
\label{fig:emission-coordinates-b}
\endgroup
\end{figure}

We start the analysis of relative source separations by examining the reconstructed emission coordinates of individual particles. The transverse projections onto the \(\tau\)--\(r_T\) and \(t\)--\(r_T\) planes are shown in Fig.~\ref{fig:emission-coordinates}, while the longitudinal projections onto the \(\tau\)--\(z\) and \(t\)--\(z\) planes are shown in Fig.~\ref{fig:emission-coordinates-b}. Here, \(\tau=\sqrt{t^2-z^2}\) denotes the proper time, \(t\) is the Cartesian time coordinate, \(r_T=\sqrt{x^2+y^2}\) is the transverse radial coordinate of a single particle, and \(z\) denotes the coordinate along the beam direction (the coordinates are taken in the laboratory frame). We present all particle species produced in the collision, except photons.

For primordial particles (first row of Fig.~\ref{fig:emission-coordinates}), the \(\tau\)--\(r_T\) distribution exhibits a clear positive correlation: particles emitted at larger transverse radii tend to appear at larger proper times. This reflects the finite space-time extension of the particlization hypersurface and the radial development of the emission region. When expressed in laboratory time, the same transverse structure extends to much larger times.
In the beam direction, shown in Fig.~\ref{fig:emission-coordinates-b}, the \(\tau\)--\(z\) distribution is approximately symmetric around \(z=0\) and displays a broad structure extending away from midrapidity, while the \(t\)--\(z\) plane shows the expected light-cone pattern associated with the relation between laboratory time and longitudinal position.

The primordial-plus-rescattering sample (second row) preserves the same qualitative structure. The close similarity between the primordial and primordial-plus-rescattering distributions indicates that hadronic rescattering does not substantially alter the global space-time geometry of the emission points for the inclusive particle sample considered here.

In contrast, the full sample including resonance decays (third row) is much more extended in all four planes, with a particularly characteristic structure appearing in the \(t\)--\(z\) projection. Although the primordial and primordial-plus-rescattering samples are concentrated close to the light-cone boundaries, the full sample also populates the interior of the cone. This reflects the fact that resonances can propagate long times before decaying. They typically move with velocities smaller than those of particles emitted directly from the particlization hypersurface or after rescattering, so their decay products are emitted at delayed times and at more moderate longitudinal positions. 
This behavior will deeply affect the long-range components in the one-dimensional source distributions, since resonance decays dominate the largest space-time separations.

\subsection{Out--Side--Long source components}
\label{sec:osl-source-components}

\begin{figure}[ht!]
\centering
\includegraphics[width=0.85\linewidth]{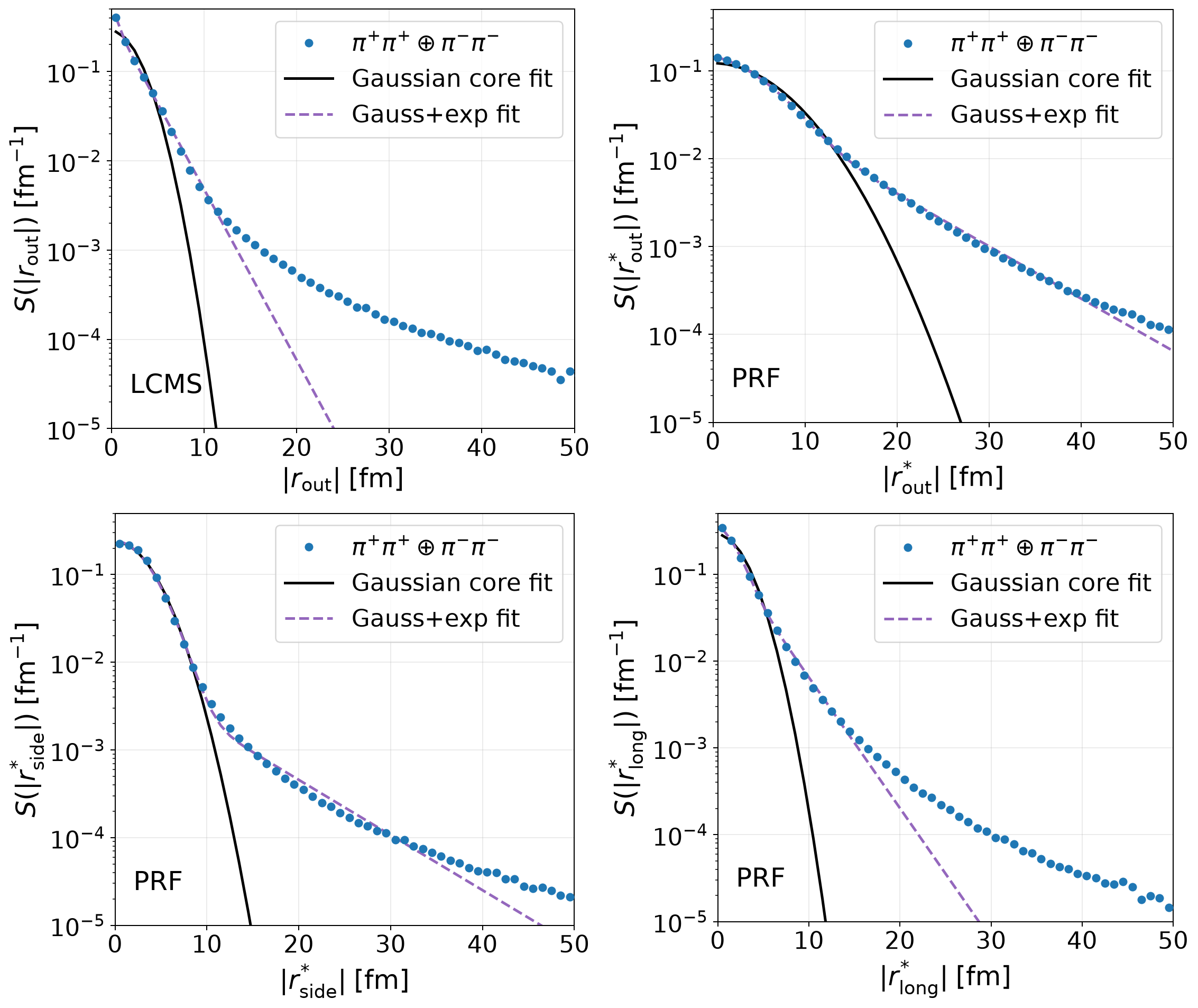}
\caption{OSL source profiles for primordial pion pairs. Upper panels show the out component in the LCMS (left) and in the PRF (right). Lower panels show the PRF side (left) and long (right) components. Solid and dashed lines show Gaussian and G+E fits, respectively.}
\label{fig:osl-primordial}
\end{figure}

\begin{figure}[!ht]
\centering
\includegraphics[width=0.85\linewidth]{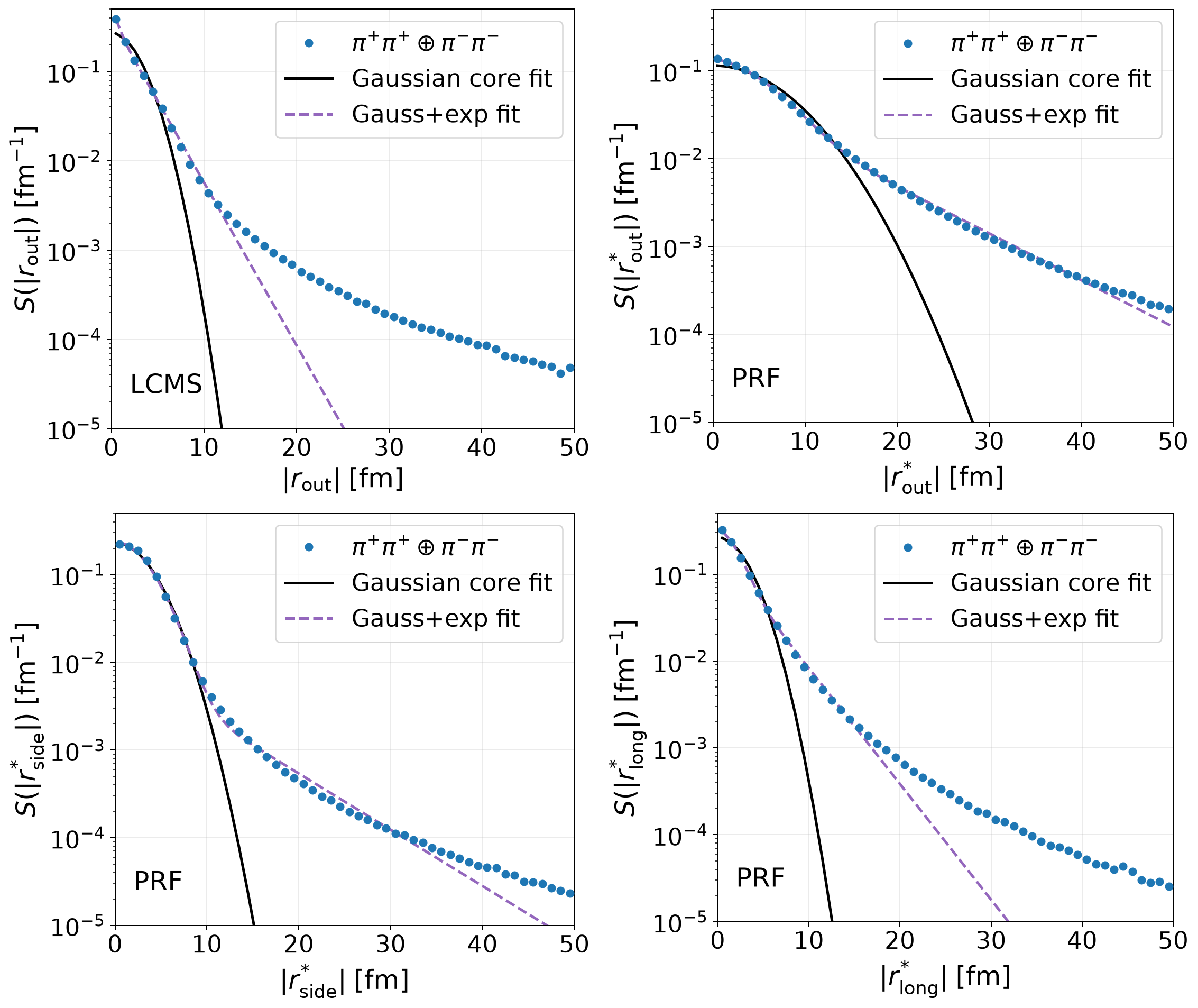}
\caption{Same as Fig.~\ref{fig:osl-primordial} but for pion pairs in the primordial-plus-rescattering selection. }
\label{fig:osl-primordial-rescattering}
\end{figure}

\begin{figure}[!htb]
\centering
\includegraphics[width=0.85\linewidth]{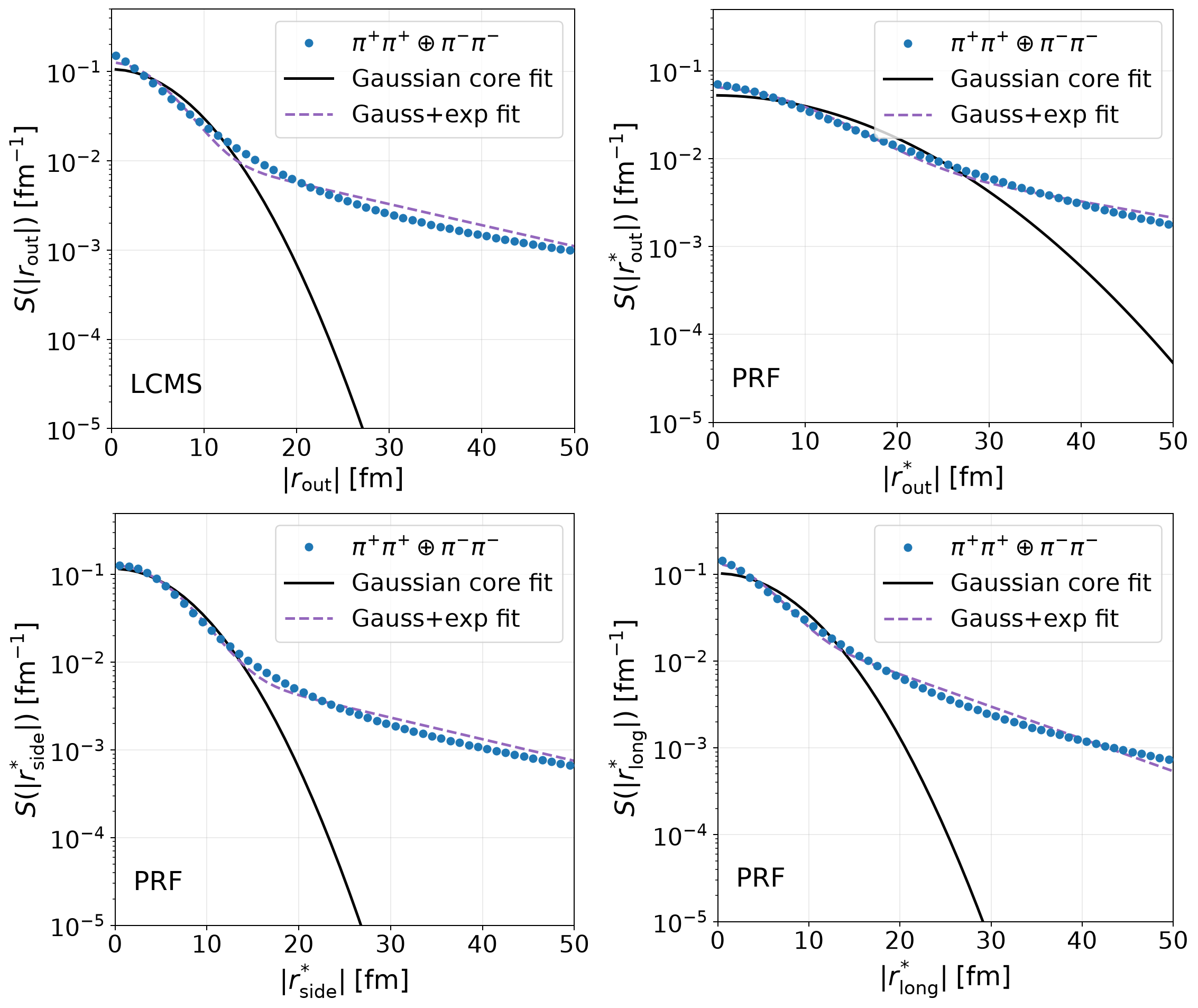}
\caption{Same as Fig.~\ref{fig:osl-primordial} but for pion pairs including primordial, rescattered, and resonance-decay contributions.}
\label{fig:osl-primordial-rescattering-decay}
\end{figure}

In this section, we start with the analysis of the source functions obtained.
We begin with the one-dimensional out--side--long source profiles, $S(r_u^*)$, which provide a component-by-component view of the spatial structure of the source. 
In Figs.~\ref{fig:osl-primordial}, ~\ref{fig:osl-primordial-rescattering} and
~\ref{fig:osl-primordial-rescattering-decay} we show the out--side--long marginal distributions for same-charge pion pairs from three different origins:
primordial particles, 
primordial-plus-rescattered particles, and  finally, all produced hadrons (all origins). 

We recall that---as shown in Eq.~\eqref{eq:lcms-to-prf-components}---under the LCMS-to-PRF boost only the out component and the emission-time difference are mixed explicitly, while the side and long components remain unchanged (which we numerically have tested). Therefore, in the upper panels we show the out distribution in both the LCMS and the PRF, to highlight how the
final frame transformation reshapes the source through the mixing of the space and time differences, and creates a more extended distribution in the PRF frame.

In Figs.~\ref{fig:osl-primordial},~\ref{fig:osl-primordial-rescattering},
~\ref{fig:osl-primordial-rescattering-decay} we also provide the result of the two fits: pure Gaussian and Gaussian plus exponential forms, and the obtained parameters, $R_u$ (for $u\in
\{\mathrm{out},\mathrm{side},\mathrm{long}\}
$) and $R_{\mathrm{core,u}}, R_{\mathrm{tail,u}}$ and $f_{\mathrm{core,u}}$, according to Eqs.~\eqref{eq:gauss} and~\eqref{eq:gaussexp} respectively, are summarized in Table~\ref{tab:osl-pion-fit-parameters}.

Let us start with primordial pion pairs in Fig.~\ref{fig:osl-primordial}. The comparison between the two parametrizations shows that non-Gaussian tails are already present at the primordial level. As we can see, the G+E form provides a better description of the large-\(r\) region. However, the exponential tail is not enough to capture pairs at large distances in the long direction. Along the out direction, the exponential tail is also not able to capture the large distance in the LCMS, but it works very well in the same direction after transformation to the PRF.

In Fig.~\ref{fig:osl-primordial-rescattering} we observe the same radii distributions when rescattered particles are added. The difference from the previous case is that the distribution of the long direction is more extended, indicating that hadronic rescattering affects more the longitudinal structure of the source. Again, the G+E form is not able to completely capture pairs at large distances for the long direction and for the out direction in the LCMS, although it perfectly does in PRF.

In Fig.~\ref{fig:osl-primordial-rescattering-decay} we include the contribution of resonance decays. The OSL source profiles show significant differences with respect to the primordial-plus-rescattering case: resonance decays generate pronounced long-range non-Gaussian tails in all OSL components, extending the distributions to much larger separations. 
In this case, we can clearly see that although the source function from all origins has a strong Gaussian core (the $f_{\mathrm{core,u}}$ weights vary between 0.7 and 0.92), the Gaussian parameterization is absolutely insufficient in the large-\(r\) region, while the G+E form describes these long-range tails very well in all directions.

\FloatBarrier

\begin{table}[h]
    \centering
    \small
    \renewcommand{\arraystretch}{0.9}
    \begin{ruledtabular}
    \begin{tabular}{lllccc}
        Pair species & Frame & Component & Primordial & Primordial + rescattering & All origins \\
        \hline
        \multicolumn{6}{c}{\(R\,[\mathrm{fm}]\)} \\
        \(\pi^+\pi^+ \oplus \pi^-\pi^-\) & LCMS & out  & $1.76$ & $1.86$ & $4.45$ \\
        & PRF  & out  & $4.38$ & $4.61$ & $9.44$ \\
         & PRF  & side & $2.32$ & $2.38$ & $4.37$ \\
         & PRF  & long & $1.84$ & $1.96$ & $4.79$ \\
        \hline
        \multicolumn{6}{c}{\(f_{\mathrm{core}}\)} \\
        \(\pi^+\pi^+ \oplus \pi^-\pi^-\) & LCMS & out  & $0.57$ & $0.59$ & $0.92$ \\
         & PRF  & out  & $0.58$ & $0.60$ & $0.73$ \\
         & PRF  & side & $0.97$ & $0.96$ & $0.90$ \\
         & PRF  & long & $0.44$ & $0.49$ & $0.70$ \\
        \hline
        \multicolumn{6}{c}{\(R_{\mathrm{core}}\,[\mathrm{fm}]\)} \\
        \(\pi^+\pi^+ \oplus \pi^-\pi^-\) & LCMS & out  & $1.95$ & $1.98$ & $4.68$ \\
         & PRF  & out  & $3.63$ & $3.68$ & $6.70$ \\
         & PRF  & side & $2.24$ & $2.28$ & $3.75$ \\
         & PRF  & long & $1.36$ & $1.40$ & $3.19$ \\
        \hline
        \multicolumn{6}{c}{\(R_{\mathrm{tail}}\,[\mathrm{fm}]\)} \\
        \(\pi^+\pi^+ \oplus \pi^-\pi^-\) & LCMS & out  & $3.30$ & $3.52$ & $26.61$ \\
         & PRF  & out  & $7.29$ & $8.16$ & $23.53$ \\
         & PRF  & side & $6.90$ & $6.77$ & $17.59$ \\
         & PRF  & long & $2.88$ & $3.26$ & $11.67$ \\
\end{tabular}
    \end{ruledtabular}
    \caption{Fit parameters for pion OSL source profiles in the LCMS and PRF. Here \(R\) is obtained from the Gaussian fit, Eq. (\ref{eq:gauss}), while \(R_{\mathrm{core}}\), \(R_{\mathrm{tail}}\), and \(f_{\mathrm{core}}\) are obtained from the G+E fit, Eq. (\ref{eq:gaussexp}).}
    \label{tab:osl-pion-fit-parameters}
\end{table}

In the results of Table~\ref{tab:osl-pion-fit-parameters} we can notice that the side projection of the source function is the one closest to a Gaussian shape (in all cases $f_{\mathrm{core}}$ is equal or above 0.90). However, we insist that such a Gaussian cannot describe alone the large-\(r\) tail.

\begin{figure}[!htb]
\centering
\includegraphics[width=\linewidth]{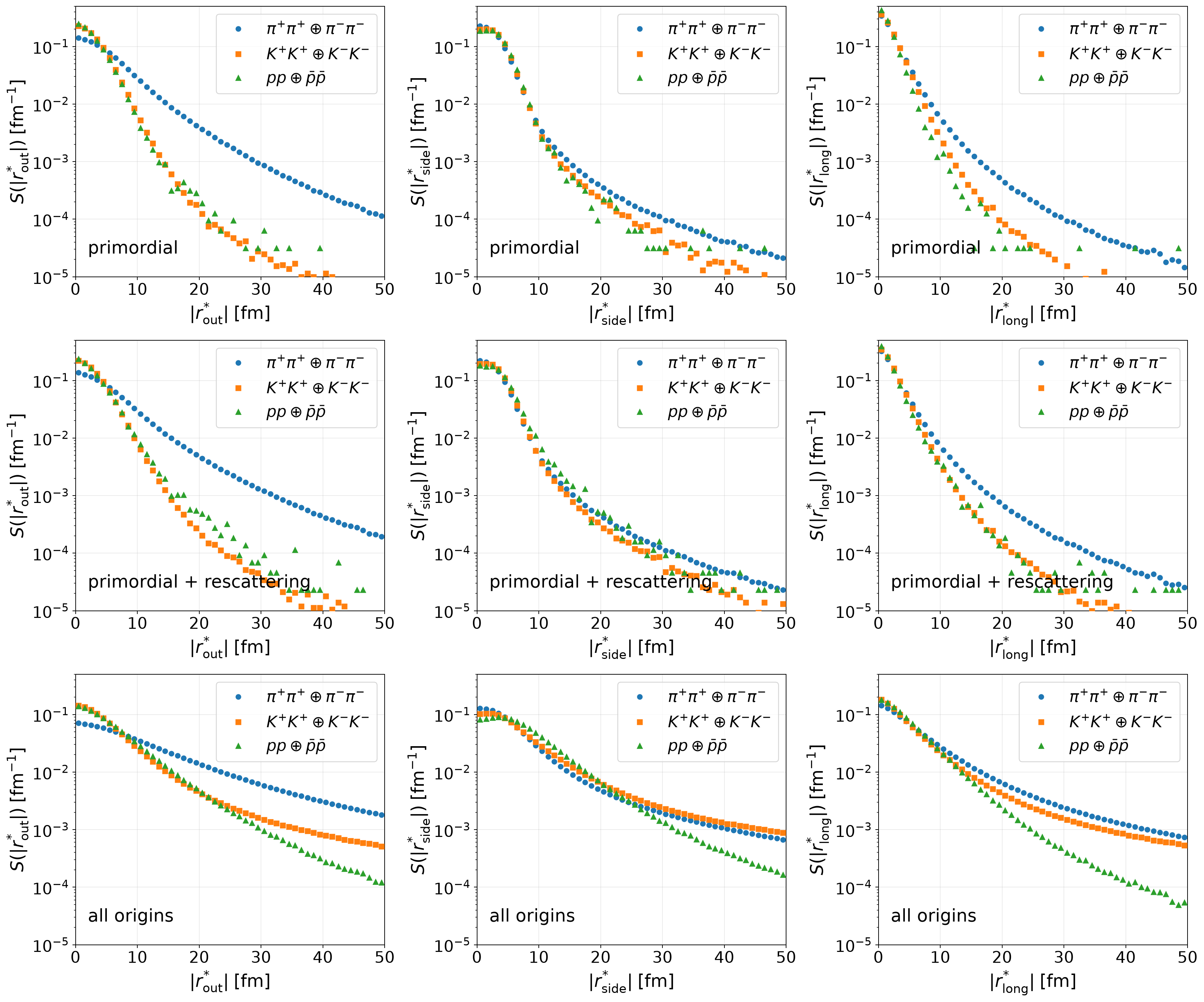}
\caption{OSL source profiles for pion (blue circles), kaon (orange squares) and proton (green triangles) pairs in PRF from three different origins:
primordial particles (upper panels), 
primordial-plus-rescattered particles (middle panels), and all origins (lower panels). }
\label{fig:osl-allspecies}
\end{figure}

We conclude that in all cases the pure Gaussian does not describe well the source function, and the 
G+E fit works much better in all cases, especially for particles from all origins, as seen in Fig.~\ref{fig:osl-primordial-rescattering-decay}.

\begin{table}[h]
    \centering
    \small
    \renewcommand{\arraystretch}{0.9}
    \begin{ruledtabular}
    \begin{tabular}{lllccc}
        Pair species & Frame & Component & Primordial & Primordial + rescattering & All origins \\
        \hline
        \multicolumn{6}{c}{\(R\,[\mathrm{fm}]\)} \\
        \(K^+K^+ \oplus K^-K^-\) & PRF & out  & $2.54$ & $2.62$ & $4.24$ \\
         & PRF & side & $2.42$ & $2.51$ & $5.23$ \\
         & PRF & long & $1.69$ & $1.78$ & $3.67$ \\
        \(pp \oplus \bar{p}\bar{p}\) & PRF & out  & $2.41$ & $2.63$ & $4.74$ \\
         & PRF & side & $2.51$ & $2.78$ & $6.08$ \\
         & PRF & long & $1.37$ & $1.53$ & $3.72$ \\
        \hline
        \multicolumn{6}{c}{\(f_{\mathrm{core}}\)} \\
        \(K^+K^+ \oplus K^-K^-\) & PRF & out  & $0.76$ & $0.74$ & $0.84$ \\
         & PRF & side & $0.99$ & $0.98$ & $0.92$ \\
         & PRF & long & $0.36$ & $0.37$ & $0.70$ \\
        \(pp \oplus \bar{p}\bar{p}\) & PRF & out  & $0.64$ & $0.51$ & $0.38$ \\
         & PRF & side & $0.99$ & $0.95$ & $0.88$ \\
         & PRF & long & $0.38$ & $0.45$ & $0.32$ \\        \hline
        \multicolumn{6}{c}{\(R_{\mathrm{core}}\,[\mathrm{fm}]\)} \\
        \(K^+K^+ \oplus K^-K^-\) & PRF & out  & $2.45$ & $2.50$ & $3.48$ \\
         & PRF & side & $2.39$ & $2.46$ & $4.53$ \\
         & PRF & long & $1.52$ & $1.53$ & $2.49$ \\
        \(pp \oplus \bar{p}\bar{p}\) & PRF & out  & $2.34$ & $2.43$ & $3.78$ \\
         & PRF & side & $2.49$ & $2.69$ & $5.70$ \\
         & PRF & long & $1.24$ & $1.18$ & $2.96$ \\
        \hline
        \multicolumn{6}{c}{\(R_{\mathrm{tail}}\,[\mathrm{fm}]\)} \\
        \(K^+K^+ \oplus K^-K^-\) & PRF & out  & $3.46$ & $3.72$ & $12.63$ \\
         & PRF & side & $7.47$ & $7.23$ & $25.70$ \\
         & PRF & long & $2.20$ & $2.40$ & $8.70$ \\
        \(pp \oplus \bar{p}\bar{p}\) & PRF & out  & $3.02$ & $3.49$ & $6.81$ \\
         & PRF & side & $8.80$ & $6.64$ & $12.53$ \\
         & PRF & long & $1.78$ & $2.34$ & $5.13$
\end{tabular}
    \end{ruledtabular}
    \caption{Fit parameters for kaon and proton OSL source profiles in the PRF. Here \(R\) is obtained from the Gaussian fit, while \(R_{\mathrm{core}}\), \(R_{\mathrm{tail}}\), and \(f_{\mathrm{core}}\) are obtained from the G+E fit.}
    \label{tab:osl-kaon-proton-fit-parameters}
\end{table}

To finish this section, we have repeated the same analysis for the other charged hadron pairs (those with the highest multiplicities). In Fig. \ref{fig:osl-allspecies} we show the kaon and proton OSL source profiles in the PRF, and in Table~  \ref{tab:osl-kaon-proton-fit-parameters} we present the results of the corresponding fits. 

For primordial particles and those with rescattering, the charged kaon and proton distributions are rather close and much narrower than the corresponding pion distributions, especially in the out direction. However, once we add the hadrons from decays, the situation changes, and the kaon source functions become closer to that of pions (in the side direction these almost coincide). Generally, the meson source function profiles appear to be wider than the proton ones. 

\FloatBarrier

\subsection{Emission-time difference distributions}
\label{sec:delta_t}

In this section, we examine the corresponding \(\Delta t\) distributions for pairs of pions of the same sign, measured in different frames.

\begin{figure}[htbp]
\centering
\includegraphics[width=\linewidth]{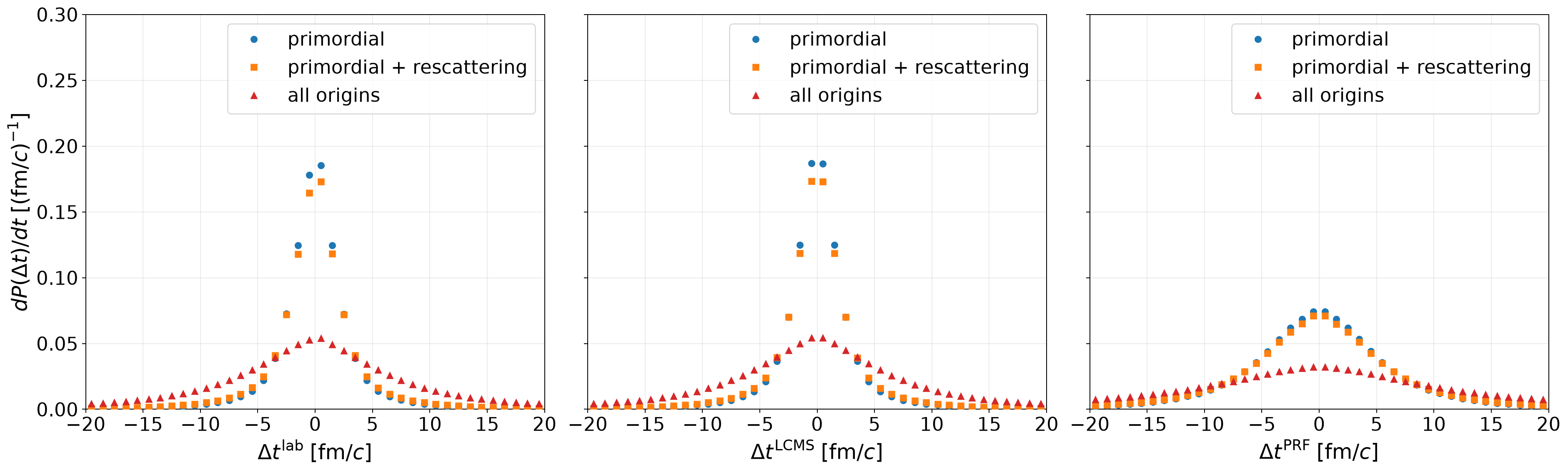}
\caption{Emission-time difference distributions for pion pairs in the laboratory frame (left), in the LCMS (middle), and in the PRF (right).}
\label{fig:dt-comparison}
\end{figure}

In Fig.~\ref{fig:dt-comparison} we present the $\Delta t$ distributions in the laboratory frame (left panel), in the LCMS (middle), and in the PRF (right). In all cases, they are peaked around \(\Delta t=0\), showing that small emission-time differences dominate the pair sample. 

To quantify these widths, Table \ref{tab:timedifferences} reports the mean time differences $\langle \Delta t \rangle$ and the standard deviations, \( \sigma_{\Delta t} = \sqrt{\langle \Delta t^2 \rangle -\langle \Delta t \rangle^2} \) in each frame and origin, with the moments of the distribution defined as
\begin{equation}
    \langle \Delta t^n \rangle =  \int_{-\infty}^{+\infty} d (\Delta t) P(\Delta t) \Delta t^n \  ,
\end{equation}
where the distribution $P (\Delta t)$ has been normalized to one. 

\begin{table}[h]
    \centering
    \begin{ruledtabular}
    \begin{tabular}{lcccccc}
        Frame & \multicolumn{2}{c}{Primordial} & \multicolumn{2}{c}{Primordial + rescattering} & \multicolumn{2}{c}{All origins} \\
        & \(\langle \Delta t \rangle\,[\mathrm{fm}/c]\) & \(\sigma_{\Delta t}\,[\mathrm{fm}/c]\) & \(\langle \Delta t \rangle\,[\mathrm{fm}/c]\) & \(\sigma_{\Delta t}\,[\mathrm{fm}/c]\) & \(\langle \Delta t \rangle\,[\mathrm{fm}/c]\) & \(\sigma_{\Delta t}\,[\mathrm{fm}/c]\) \\
        \hline
        Lab frame & \(0.01\) & \(6.92\) & \(0.00\) & \(7.59\) & \(-0.07\) & \(361.56\) \\
        LCMS & \(0.00\) & \(6.80\) & \(0.00\) & \(7.50\) & \(-0.06\) & \(336.01\) \\
        PRF & \(0.00\) & \(9.71\) & \(0.00\) & \(11.10\) & \(-0.05\) & \(331.54\) \\
    \end{tabular}
    \end{ruledtabular}
    \caption{Mean values \(\langle \Delta t \rangle\) and standard deviations \(\sigma_{\Delta t}\) of the emission-time difference distributions for the three reference frames and particle-origin selections. The ``all origins'' selection includes primordial, rescattered, and resonance-decay contributions.}
    \label{tab:timedifferences}
\end{table}

In terms of particle-origin selections, primordial and primordial-plus-rescattering samples remain relatively close to each other, whereas the  all-origins sample produces a much broader distribution, with an enhanced weight at large \(|\Delta t|\), reflecting the delayed emission associated with resonance decays. 

In the laboratory frame and in the LCMS, the primordial and primordial-plus-rescattering distributions are similar, and sharply peaked around \(\Delta t=0\) with a rapid fall. By contrast, the distribution including resonance decays is broader and smoother, 
with a very large standard deviation, $\ge 300$ fm/c, suggesting that it has a long-tail structure. 

Note that in experimental studies, 
the $\Delta t$ dependence of the pair source function is usually integrated over~\cite{Heinz:1999rw,Lisa:2005dd}, as we did in the previous (and future) sections. However, in simulations, the emission time difference distribution can be addressed, and, in particular, we can compare our primordial and primordial-plus-rescattering PRF distributions with the results presented in Ref.~\cite{Kisiel:2025jbg}, where the authors obtain a qualitatively similar distribution (see their Fig. 10). 

\subsection{Radial source functions}

In this section, we construct the one-dimensional radial $r_{\mathrm{full}}^{*}$ source distributions according to Eq.~\eqref{eq:radial-separation}.  This allows us to quantify the overall size of the emission source and to characterize possible long-range contributions beyond the compact core.

This reduction of the number of dimensions, yielding a spherically averaged source, is sometimes performed in experimental data when the statistics are low for some pairs of hadrons, and a three-dimensional analysis would generate too much uncertainty.

\begin{figure}[htbp]
\centering
\includegraphics[width=\linewidth]{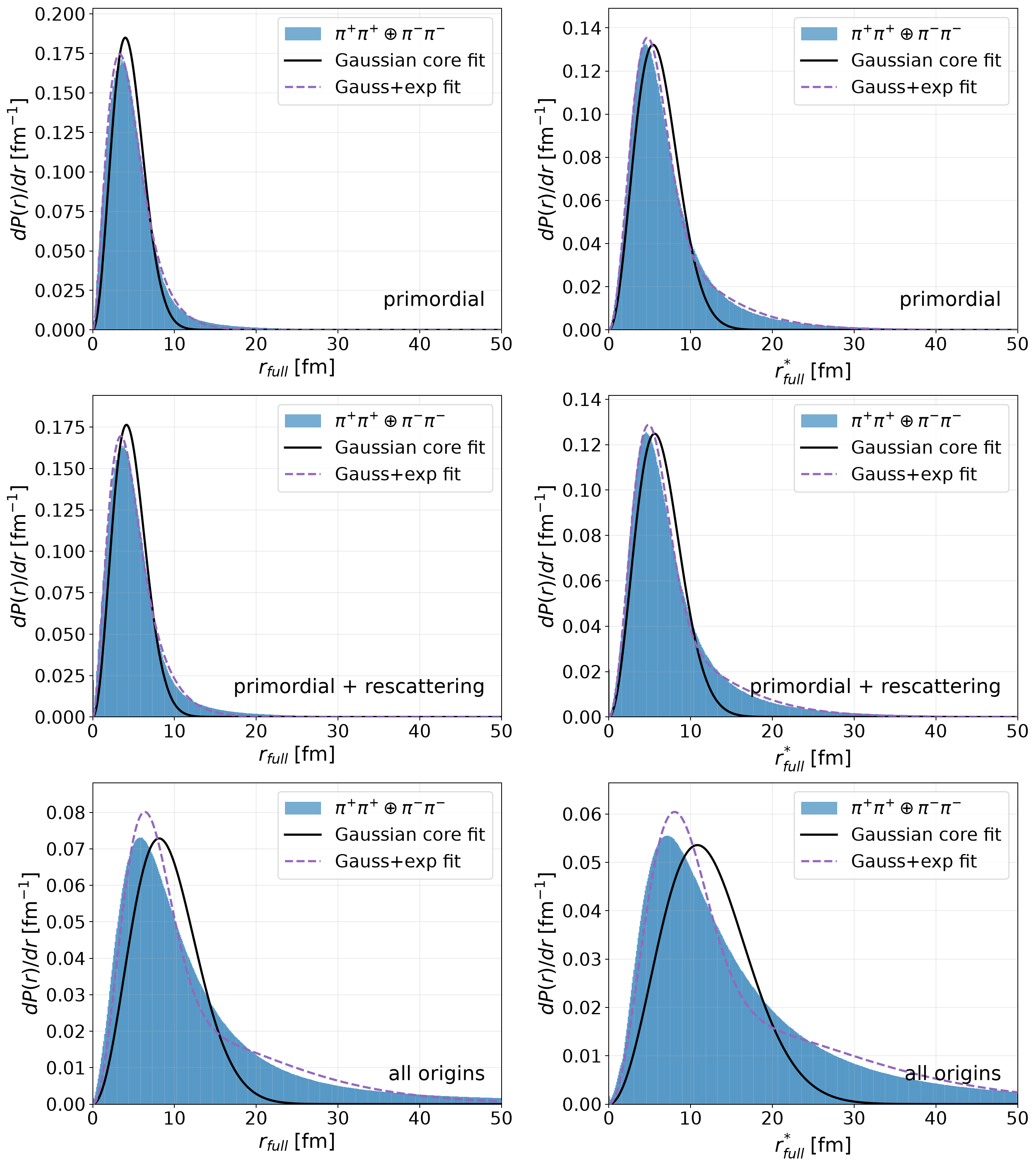}
\caption{Radial source distributions for the pion sample. Rows correspond to primordial particles, primordial-plus-rescattered particles, and the full sample including resonance-decay contributions, respectively. The left column shows the LCMS distributions, while the right column shows the corresponding PRF distributions.}
\label{fig:radial-sources}
\end{figure}

Figure~\ref{fig:radial-sources} compares the one-dimensional radial source distribution ($P(r^*_{\mathrm{full}})=r_{\mathrm{full}}^{*2} S(r_{\mathrm{full}}^{*})$) for charged pion pairs in LCMS (left column) and PRF (right column) for the primordial (upper row), primordial-plus-rescattering (middle row) and the full sample (lower row) selection. Table~\ref{tab:radial-radii} reports the results of the pure Gaussian and the G+E fits. 
 
We can see that the primordial and primordial-plus-rescattering distributions are dominated by a compact core, with only a small tail, 
although again we see that the pure Gaussian fit does not describe well the data. 
The full sample, on the other hand, develops a much more pronounced non-Gaussian tail, as can be expected from the results of Sec.~\ref{FO_distribution}. The LCMS-to-PRF boost systematically leads to a spatially more extended source.  

Such long-range, non-Gaussian structures in pion source functions have been observed experimentally in source-imaging analyses at RHIC and SPS~\cite{PHENIX:2007LongRange,PHENIX:2008SourceBreakup,Chung:2007SPS}. Similar deviations from a Gaussian shape at large relative separations have also been reported for charged-kaon source functions at RHIC~\cite{PHENIX:2009KaonSource}.

\begin{table}[h]
    \centering
    \begin{ruledtabular}
    \begin{tabular}{llccc}
        Pair species & Frame & Primordial & Primordial + rescattering & All origins \\
        \hline
        \multicolumn{5}{c}{\(R\,[\mathrm{fm}]\)} \\
        \(\pi^+\pi^+ \oplus \pi^-\pi^-\) & LCMS & $1.99$ & $2.07$ & $4.06$ \\
        \(\pi^+\pi^+ \oplus \pi^-\pi^-\) & PRF  & $2.70$ & $2.79$ & $5.42$ \\
         \hline
        \multicolumn{5}{c}{\(f_{\mathrm{core}}\)} \\
        \(\pi^+\pi^+ \oplus \pi^-\pi^-\) & LCMS & \(0.07\) & \(0.15\) & \(0.86\) \\
        \(\pi^+\pi^+ \oplus \pi^-\pi^-\) & PRF  & \(0.77\) & \(0.80\) & \(0.85\) \\
        \hline
        \multicolumn{5}{c}{\(R_{\mathrm{core}}\,[\mathrm{fm}]\)} \\
        \(\pi^+\pi^+ \oplus \pi^-\pi^-\) & LCMS & $1.77$ & $1.69$ & $3.05$ \\
        \(\pi^+\pi^+ \oplus \pi^-\pi^-\) & PRF  & $2.27$ & $2.33$ & $3.82$ \\
        \hline
        \multicolumn{5}{c}{\(R_{\mathrm{tail}}\,[\mathrm{fm}]\)} \\
        \(\pi^+\pi^+ \oplus \pi^-\pi^-\) & LCMS & $1.59$ & $1.74$ & $6.29$ \\
        \(\pi^+\pi^+ \oplus \pi^-\pi^-\) & PRF  & $3.52$ & $3.86$ & $8.25$ \\
    \end{tabular}
    \end{ruledtabular}
    \caption{
    Fit parameters for radial source distributions for charged pion pairs in LCMS and PRF. Here \(R\) is obtained from the Gaussian fit, while \(R_{\mathrm{core}}\), \(R_{\mathrm{tail}}\), and \(f_{\mathrm{core}}\) are obtained from the G+E fit.}
    \label{tab:radial-radii}
\end{table}

\begin{figure}[htbp]
\centering
\includegraphics[width=\linewidth]{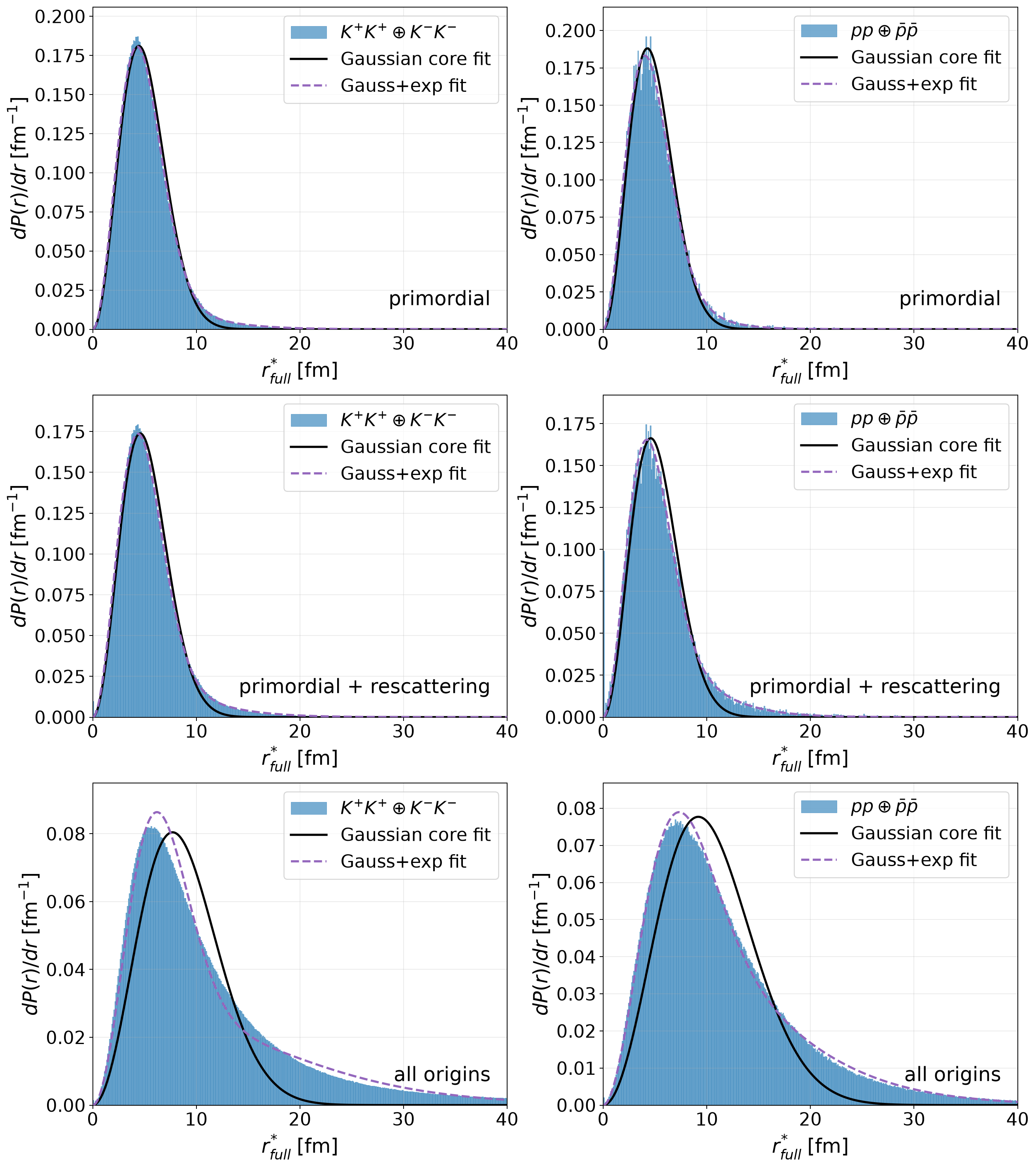}
\caption{Radial source distributions in the PRF for kaon and proton pairs. Rows correspond to primordial particles, primordial-plus-rescattered particles, and the full sample including resonance-decay contributions, respectively. The left column shows kaon pairs and the right column proton pairs.}
\label{fig:radial-kaon-proton}
\end{figure}

To complete this section, we add similar radial source distributions for kaon and proton pairs (only in PRF) in Fig.~\ref{fig:radial-kaon-proton}. The resulting parameters of the fits appear in Table~\ref{tab:radial-radii-kaon-proton}. We can see that the kaon and proton radial distributions are rather similar and a bit narrower than the pion ones. However, for all three studied particle pairs, we observe the same qualitative picture.  

The Gaussian and G+E fits provide a quantitative way to separate the compact core from the long-range contribution. For the primordial and primordial-plus-rescattering selections, the G+E form gives an accurate description of both the dominant core and the small residual tail, while the purely Gaussian fit mainly captures the compact part of the distribution. 

When resonance-decay contributions are included, the long-range component becomes much more pronounced. In this case, the G+E parametrization still improves substantially over a single Gaussian, but the full detailed tail description would likely require a more complicated ansatz with more fitting parameters. Thus, the radial source distribution of all produced particles appears to be technically more complicated than the corresponding OSL source profiles.

\begin{table}[htbp]
\centering
\begin{ruledtabular}
\begin{tabular}{llccc}
Pair species & Frame & Primordial & Primordial + rescattering & All origins \\
\hline
\multicolumn{5}{c}{\(R\,[\mathrm{fm}]\)} \\
\(K^+K^+ \oplus K^-K^-\) & PRF  & $2.23$ & $2.29$ & $3.85$ \\
\(pp \oplus \bar{p}\bar{p}\) & PRF  & $2.13$ & $2.30$ & $4.59$ \\
\hline
\multicolumn{5}{c}{\(f_{\mathrm{core}}\)} \\
\(K^+K^+ \oplus K^-K^-\) & PRF  & \(0.83\) & \(0.81\) & \(0.82\) \\
\(pp \oplus \bar{p}\bar{p}\) & PRF  & \(0.47\) & \(0.62\) & \(0.45\) \\
\hline
\multicolumn{5}{c}{\(R_{\mathrm{core}}\,[\mathrm{fm}]\)} \\
\(K^+K^+ \oplus K^-K^-\) & PRF  & $2.13$ & $2.16$ & $2.96$ \\
\(pp \oplus \bar{p}\bar{p}\) & PRF  & $2.05$ & $2.07$ & $3.40$ \\
\hline
\multicolumn{5}{c}{\(R_{\mathrm{tail}}\,[\mathrm{fm}]\)} \\
\(K^+K^+ \oplus K^-K^-\) & PRF  & $2.49$ & $2.63$ & $5.60$ \\
\(pp \oplus \bar{p}\bar{p}\) & PRF  & $1.83$ & $2.41$ & $4.61$ \\
\end{tabular}
\end{ruledtabular}
\caption{Fit parameters for kaon and proton 
radial source distributions in PRF. Here \(R\) is obtained from the Gaussian fit, while \(R_{\mathrm{core}}\), \(R_{\mathrm{tail}}\), and \(f_{\mathrm{core}}\) are obtained from the G+E fit.}
\label{tab:radial-radii-kaon-proton}
\end{table}

\subsection{\texorpdfstring{\(m_T\)}{mT} dependence of the core radius}

In this section, we finally present the transverse mass dependence of \(R_{\mathrm{core}}\), defined as the radius of the Gaussian component in the G+E fit of the PRF radial source distributions. This quantity provides a representative scale of the dominant emission region, which is sensitive to the collective expansion of the system~\cite{Wiedemann:1995au,Wiedemann:1999qn,ALICE:2015PionKaonProton}. As a consequence, a larger $m_T$ generally selects a smaller effective emission region. This analysis is motivated by recent ALICE studies of common hadron-emission sources and approximate \(m_T\)-scaling in small systems~\cite{ALICE:2023sjd}.

\begin{figure}[htbp]
\centering
\includegraphics[width=0.75\linewidth]{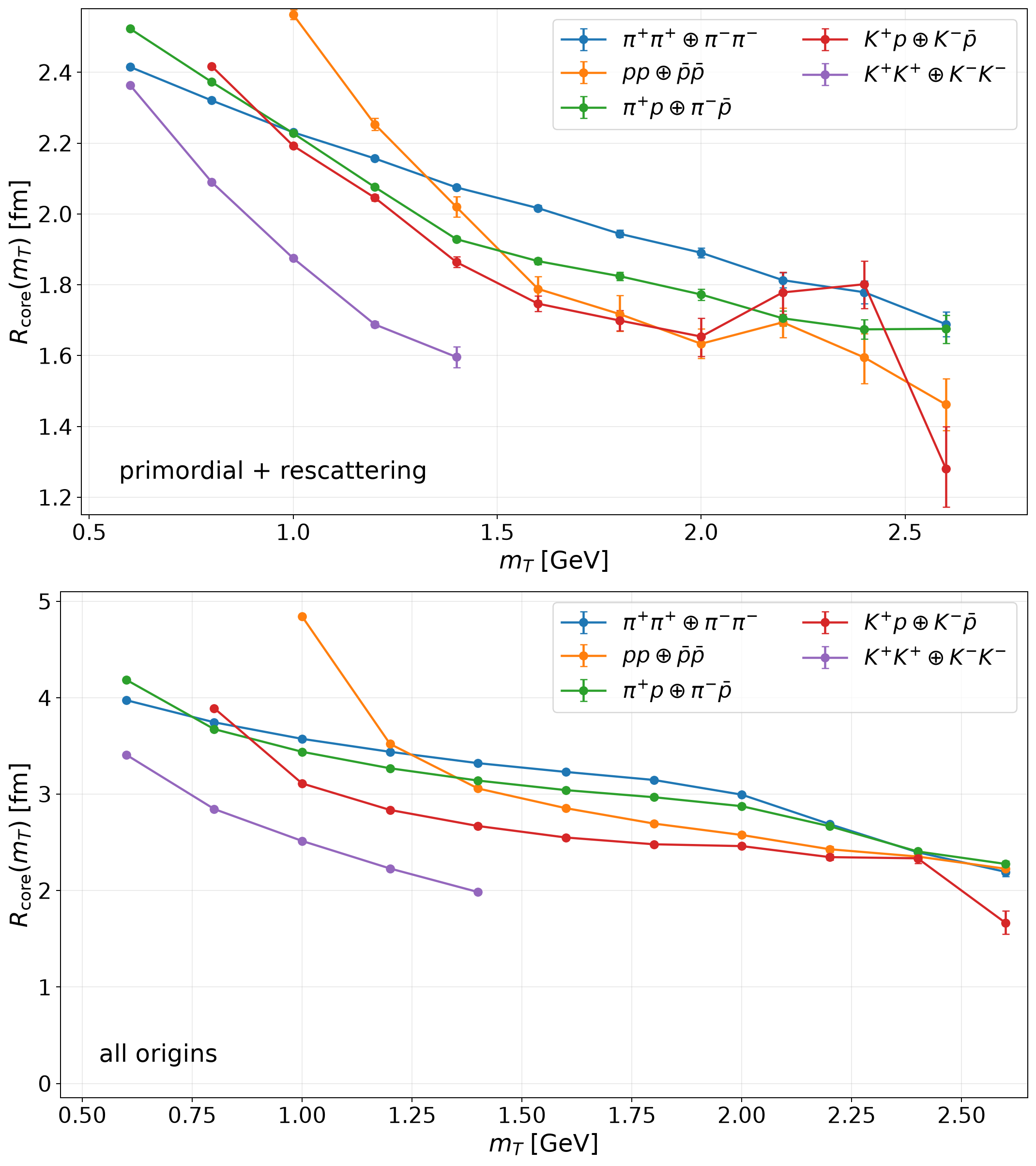}
\caption{Transverse-mass dependence of the Gaussian core radius $R_{\mathrm{core}}$, extracted from the PRF radial source distributions using the Gaussian component of the G+E fit. The upper panel corresponds to the primordial-plus-rescattering selection, while the lower panel shows the full sample including primordial, rescattered, and resonance-decay contributions. Both panels include same-species and cross-species pair combinations.}
\label{fig:rcore-mt}
\end{figure}

Figure~\ref{fig:rcore-mt} shows \(R_{\mathrm{core}}(m_T)\) for the statistically most significant pair combinations: \(\pi^+\pi^+ \oplus \pi^-\pi^-\), \(pp \oplus \bar p\bar p\), \(\pi^+p \oplus \pi^-\bar p\), \(K^+p \oplus K^-\bar p\), and \(K^+K^+ \oplus K^-K^-\).  To see the effect of the large non-Gaussian contribution generated by resonance decays, we show in the 
 upper panel $R_{\mathrm{core}}$ that corresponds to the primordial-plus-rescattering selection, while in the lower panel we show $R_{\mathrm{core}}$ obtained the full sample analysis,  including primordial, rescattered, and resonance-decay contributions.

Comparing the extracted core radii with those from the analysis of pp collisions at LHC,  Ref.~\cite{Fabbietti:2020bfg}, we can see a similar trend of decrease with increasing \(m_T\). Our radii are somewhat larger, which is consistent with the larger system size.

Within the present uncertainties, the different combinations of the same and different species follow a common approximate trend, suggesting that the compact part of the emission source is largely governed by a common \(m_T\)-dependent scale. For larger \(m_T\), the uncertainties increase, particularly for the less abundant pair combinations, and therefore the apparent deviations from a common trend should be interpreted with caution.

Our results for very peripheral collisions can also be compared to data for peripheral $\text{Pb}-\text{Pb}$ collisions at LHC. For example, Fig. 8 from  Ref.~\cite{ALICE:2015PionKaonProton} is very similar to our Fig.~\ref{fig:rcore-mt}, and the results are quite consistent, considering that we study more peripheral collisions and, according to the clear trend, our radii should be lower than theirs. Furthermore, this figure clearly shows that as we move to more central $\text{Pb}-\text{Pb}$ collisions, $m_T$-scaling breaks down even further. It actually motivates our decision to concentrate on very peripheral collisions. 
Another recent paper from ALICE~\cite{ALICE:2025wuy} concentrates on proton-proton femtoscopy, and the obtained radii for 30-50$\%$ of centrality are close to ours. 

\section{Conclusions~\label{sec:conclusions}}

In this work, we performed a comprehensive microscopic analysis of the space-time femtoscopic emission source in low-multiplicity ($80$--$90\%$ centrality) $\text{Pb}-\text{Pb}$ collisions at $\sqrt{s_{NN}} = 5.02~\text{TeV}$ using a SMASH--vHLLE--SMASH hybrid transport approach. By using the full phase-space particle histories recorded in the hadronic afterburner, we reconstructed the emission points of individual hadrons, allowing us to explicitly isolate and compare the contributions from primordial, rescattered, and hadrons from resonance decays.

We found a sizable differences between the effect coming from particles suffering hadronic rescattering and those coming from resonance decays. While rescattering produces only minor spatial modifications to the primary emission geometry, the resonance decays introduce significant time delays, extending the space-time emission region into the light-cone interior and populating the long-range non-Gaussian tails across all OSL and radial source components.

By comparing source distributions in the LCMS and in the PRF reveals that the Lorentz boost from LCMS to PRF strongly couples the emission-time difference $\Delta t$ to the out component, visibly expanding the spatial profile along $r_{\text{out}}^*$.

We have clearly determined that, for all examined particle pairs ($\pi\pi$, $KK$, $pp$, and cross-species), the pure Gaussian parameterization does not describe well the extended source structure, particularly when resonance contributions are included. A combined Gaussian plus exponential form, $S_{\text{G+E}}(r^*)$, provides a phenomenological separation between a compact Gaussian emission core ($R_{\text{core}}$) and a long-range exponential component ($R_{\text{tail}}$).

Finally, the extracted Gaussian core radii $R_{\text{core}}(m_T)$ in the PRF exhibit a clear monotonic decrease with increasing transverse mass $m_T$, consistent with the results in pp collisions analyzed by the ALICE collaboration. Moreover, both same-species and cross-species combinations lie approximately on a common trend, mirroring the $m_T$-scaling behavior typically observed in pp collisions. 
The quantitative values of $R_{\text{core}}$ are also of the same order, but somewhat larger, which can be naturally explained from the difference of the colliding systems. This might indicate that in peripheral $\text{Pb}-\text{Pb}$ collisions, the compact core of particle freeze-out is also dictated by common hydrodynamic flow and kinematic expansion effects.

These results suggest that future high-precision experimental femtoscopy and source-imaging analyses need to adopt two-component parameterizations beyond the single Gaussian approximation to accurately dissect the underlying source geometry.

\begin{acknowledgments}

We acknowledge Laura Serksnyte for general discussions and for bringing relevant references on femtoscopy to our attention.

This work has been supported by the project numbers CEX2019-000918-M (Unidad de Excelencia ``Mar\'ia de Maeztu'') and PID2023-147112NB-C21, financed by the Spanish MCIN/ AEI/10.13039/501100011033/; and by Contract 2021 SGR 171 by the Generalitat de Catalunya.
JMT-R also thanks Grant No. 402942/2024 by the Brazilian CNPq (National Council for Scientific and Technological
Development).
\end{acknowledgments}

\bibliography{references}

\end{document}